# Coherent Acoustic Phonons in Plasmonic Nanoparticles: Elastic Properties and Dissipation at Low Temperatures

Hilario D. Boggiano,[1] Thomas Possmayer,[2] Luis Morguet,[2] Lin Nan,[2] Luca Sortino,[2] Stefan A. Maier,[3,4] Emiliano Cortés,[2] Gustavo Grinblat,[1,5] Andrea V. Bragas,*[,1,5] Leonardo de S. Menezes*[,2,6]

[1] Universidad de Buenos Aires, Facultad de Ciencias Exactas y Naturales, Departamento de Física. 1428 Buenos Aires, Argentina.
[2] Chair in Hybrid Nanosystems, Nanoinstitute Munich, Faculty of Physics, Ludwig-Maximilians-Universität München, 80539 München, Germany.
[3] School of Physics and Astronomy, Monash University, Clayton, Victoria 3800, Australia.
[4] Department of Physics, Imperial College London, London SW7 2AZ, U.K.
[5] CONICET - Universidad de Buenos Aires, Instituto de Física de Buenos Aires (IFIBA). 1428 Buenos Aires, Argentina.
[6] Departamento de Física, Universidade Federal de Pernambuco, 50670-901 Recife-PE, Brazil.

* Email for A.V.B.: bragas@df.uba.ar. Email for L. de S.M.: L.Menezes@physik.uni-muenchen.de

## Abstract

We studied the frequency and quality factor of mechanical plasmonic nanoresonators as a function of temperature, ranging from ambient to 4 K. Our investigation focused on individual gold nanorods and nanodisks of various sizes. We observed that oscillation frequencies increase linearly as temperature decreases until saturation is reached at cryogenic temperatures. This behavior is explained by the temperature dependence of the elastic modulus, with a Debye temperature compatible with reported bulk values for gold. To describe the behavior of the quality factor, we developed a model considering the nanostructures as anelastic solids, identifying a dissipation peak around 150 K due to a thermally activated process, likely of the Niblett-Wilks mechanism type. Importantly, our findings suggest that external dissipation factors are more critical to improving quality factors than internal friction, which can be increased by modifying the nanoresonator's environment. Our results enable the design of structures with high vibration frequencies and quality factors by effectively controlling external losses.

## Introduction

Mechanical nanoresonators play a pivotal role in various cutting-edge scientific and technological applications. In the realm of quantum technologies, they hold particular promise. Because of their minuscule size, nanoresonators can sustain extremely high-frequency (high-$f$) vibrational modes up to tens of GHz.[1–4] Consequently, the ground state energy of these systems can surpass thermal energy, which would enable the observation of quantum phenomena even at room temperature.[5–7] However, internal and external loss mechanisms limit these nanoresonators' lifetimes, which typically shorten with higher frequencies. As the quality factor $Q$ dictates how quickly energy dissipates, the quantity $f \times Q$ serves as an effective merit factor for describing the performance and sensitivity of nanoresonators. In cavity optomechanics, a critical condition for neglecting thermal decoherence over one mechanical period is $f \times Q > k_B T/h$,[8] where $h$, $k_B$, and $T$ are Planck's and Boltzmann's constants, and temperature, respectively. For room temperature, this condition translates into $f \times Q > 6 \times 10^{12}$ Hz.[9,10] Proposals of suitable nanoresonators with high-$f$ and high-$Q$, needed for observing room temperature quantum effects as well as for ultrasensitive sensing[11,12] are still an open challenge in the community.

Optically launched coherent phonons in plasmonic nanoparticles have been widely used as a route to make nano-objects vibrate at very high-$f$, typically in the range of a few to tens of GHz. Improvements in nanofabrication have made it possible to design a wide variety of them, which offer both sensitivity in optical response throughout the visible range and high tunability in mechanical response at GHz.[2,13] The operation of these plasmonic nanoresonators lies in the resonant excitation of a large population of hot electrons with a pulse of light, which eventually produces a coherent motion of the lattice ions upon de-excitation.[14] This minute motion can be read with a second pulse of light since its transmission or reflection through the sample is modified with the nanoparticle's motion, both by a change of shape and the effect of the deformation potential.[15] Typically, this experiment is done in a pump-probe configuration, where the temporal dynamics is reconstructed by delaying the probe to the pump step by step and recording the differential transmission or reflection in each point. The sensitivity of these experiments allows for the monitoring of individual nanoresonators.[16–18]

These nanoresonators generally show very modest $Q$ values of no more than a few tens throughout the available bibliography on the field. However, in recent work, it is reported that gold nanoplates supported by low-density substrates have high-$Q$ values reaching around 200 at room temperature.[1] Strong coupling between vibrational modes could be demonstrated using these systems, which is of great interest for the quantum technologies mentioned above. However, the other side of the coin of these mechanical plasmonic nanoresonators is that they can be coupled to the substrate as a mechanism to dissipate the thermoelastic energy generated optically. While this lowers the $Q$, it allows these nanoresonators to become sources of surface acoustic waves (SAWs) generated entirely optically,[19] an effect of technological relevance for information processing in photonic platforms with increasingly challenging requirements of speed and miniaturization. In addition, releasing energy to the substrate allows the coupling of the different parts in nanoresonator arrangements[20,21] or between different acoustic modes of the same nano-object,[22] or the focusing of hypersonic acoustic energy.[23]

Several mechanisms contribute to the energy dissipation of an oscillator connected to the environment, which can be divided into internal and external causes. In metals, thermoelastic damping is identified as one of the most critical internal dissipation mechanisms, which refers to the process by which the mechanical energy of a resonator is irreversibly converted into thermal energy. This attenuation occurs due to the dissipation of the energy produced when heat flows from the hot compressed regions to the cooler expanded regions. Additionally, intrinsic material losses due to crystal defects (*e.g.,* dislocations and grain boundaries) also limit the quality factor of acoustic vibrations. They are the dominant damping channel in polycrystalline structures and amorphous systems, as obtained by lithographic fabrication.[24,25] Regarding the external sources, damping due to acoustic energy transfer into the environment arises from mechanical contact with the substrate. Its magnitude is determined by the impedance mismatch between nanoparticle and substrate materials and, in addition, the quality of the mechanical contact.[26] Ultimately, among these mechanisms, one seeks to distinguish between those that fundamentally set an upper limit on the quality factor and those that could be eliminated through improved design and fabrication.

Despite the large volume of discussion in the literature published so far, where nanoresonators' frequency and dissipation mechanisms are analyzed, there are no measurements other than at room temperature.[21,27,28] To the best of our knowledge, the only exception is a paper by Wang et al.,[29] where the oscillator frequency is monitored at low temperatures, but solely to indirectly characterize the surrounding material mechanical behaviour. Elastic properties of solids are determined by microscopic interactions and the resulting lattice dynamics, which are known to be sensitive to temperature. It is, therefore, important to understand the influence of temperature on vibrational frequencies and lifetimes in nanomechanical resonators to unravel the nature of the underlying mechanisms. In this work, we present the study of the frequency and quality factor of plasmonic nanoresonators as a function of temperature in the range from 4 K to room temperature. This study not only provides a comprehensive picture of the behavior of nanoresonators to determine possible loss mechanisms but also enables new tools to understand what happens to these nanoresonators when they are, for instance, part of hybrid systems designed to operate at low temperatures.

**Results and Discussions**

Gold nanorods and nanodisks were fabricated on borosilicate substrates using electron beam lithography. The design of the fabricated nanostructures is shown in Figure 1a, together with representative scanning electron microscope (SEM) images in Figure 1b (see fabrication details in Methods section). The nanorods have a nominal length ranging from $L$ = 220 nm to $L$ = 280 nm (in 10 nm steps), width $W$ = 100 nm, and height $H_r$ = 45 nm, and are attached to the substrate via a 2 nm thick Cr adhesion layer. On the other hand, the gold nanodisks have a nominal diameter ranging from $D$ = 100 nm up to $D$ = 245 nm (100, 140, 200, and 245 nm), and height $H_d$ = 35 nm, and were fabricated with a molecular adhesion layer (MPTMS). Figure 1c presents the simulated absorption cross-section of the nanostructures.

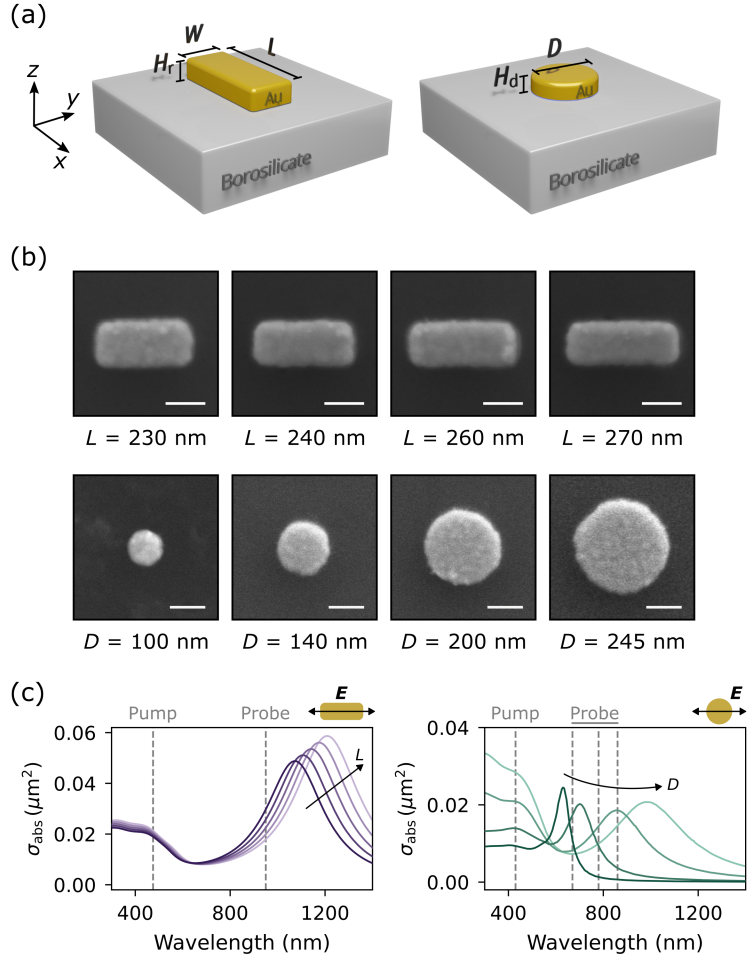

**Figure 1**. Plasmonic nanostructures and their optical response. (a) Schematic view of the fabricated plasmonic gold nanoantennas. (Left) Nanorods with nominal lengths ranging from $L = 230$ nm to 270 nm (in 10 nm steps), width $W = 100$ nm, and height $H_r = 45$ nm, attached to a borosilicate substrate via a 2 nm Cr adhesion layer. (Right) Nanodisks with diameters ranging from $D = 100$ nm to 245 nm (100, 140, 200, and 245 nm), and height $H_d = 35$ nm, coupled to the substrate with a molecular adhesion layer (MPTMS). (b) Scanning electron microscope (SEM) images of the fabricated nanostructures. Scale bars: 100 nm. (c) Calculated absorption cross-sections, $\sigma_{abs}$, for rods (left) and disks (right) of different sizes. The arrow at the top right of each panel indicates the direction of incident polarization. In both panels, the vertical dashed lines mark the peak wavelengths of the pump and probe beams.

Single-particle differential reflection measurements were conducted inside a closed-cycle cryostat (Montana Instruments Cryostation s50) with a collinear two-color pump-probe setup, as illustrated in Figure 2a. A Ti:sapphire laser with 80 MHz repetition rate and pulse duration of 140 fs was used to pump an optical parametric oscillator (OPO) at 860 nm wavelength. The frequency-doubled fundamental beam (at 430 nm wavelength) was chopped at 1.9 kHz and utilized to pump the individual nanostructures. For the probe beam, the frequency-doubled OPO signal was tuned to the red side of the plasmon resonance for 100 nm disks (670 nm) and to the blue side for 140 nm and 200 nm disks (670 and 780 nm wavelength, respectively), based on simulated absorption spectra shown in Figure 1c (right panel). The 245 nm disks were probed with the fundamental laser output at 860 nm wavelength. For the nanorods, the Ti:sapphire laser output was tuned to 950 nm for the probe beam and its second harmonic at 475 nm for the pump beam, regardless of the nanoparticle length, as indicated in Figure 1c (left panel). After being focused onto the sample mounted within the cryostat, the reflected light was optically filtered, and the remaining probe beam was directed to a photoreceiver. The modulated transient reflection signal was then extracted by a lock-in amplifier synchronized with the optical chopper.

Figure 2b,c (top panels) shows the measurement of the differential reflection ($\Delta R/R$) trace of a gold nanorod with length $L$ = 250 nm (Figure 2b) and a gold nanodisk with diameter $D$ = 140 nm (Figure 2c), both at room temperature ($T$ = 300 K). The decaying background arising from the cooling of the nanoparticle was fitted using a single exponential curve and subtracted from the data, leaving only the phononic-oscillatory component (Figure 2b,c, bottom panels). One or more vibrational modes were detected, depending on the nanostructure. This behavior was accounted for by using a sum of exponentially damped sinusoids to model the signals:

$$\Delta R/R = \sum_n A_n e^{-t/\tau_n} \cos(2\pi f_n + \phi_n) \qquad (1)$$

where the differential reflection is described by $n$ modes of amplitude $A_n$, frequency $f_n$, damping time $\tau_n$, and phase $\phi_n$. The quality factor $Q$ is calculated as $Q_n = \pi f_n \tau_n$. The fast Fourier transform (FFT) of the phonon signals displayed in the bottom panels of Figure 2b,c reveals a single oscillation mode for the nanorod (Figure 2d), with a frequency of about 5.5 GHz (labeled r1) and three different vibrational modes for the nanodisk (Figure 2e), the main one with a frequency of about 10 GHz (d2), and other two with lower amplitudes and frequencies of about 7.4 GHz (d1) and 11.6 GHz (d3).

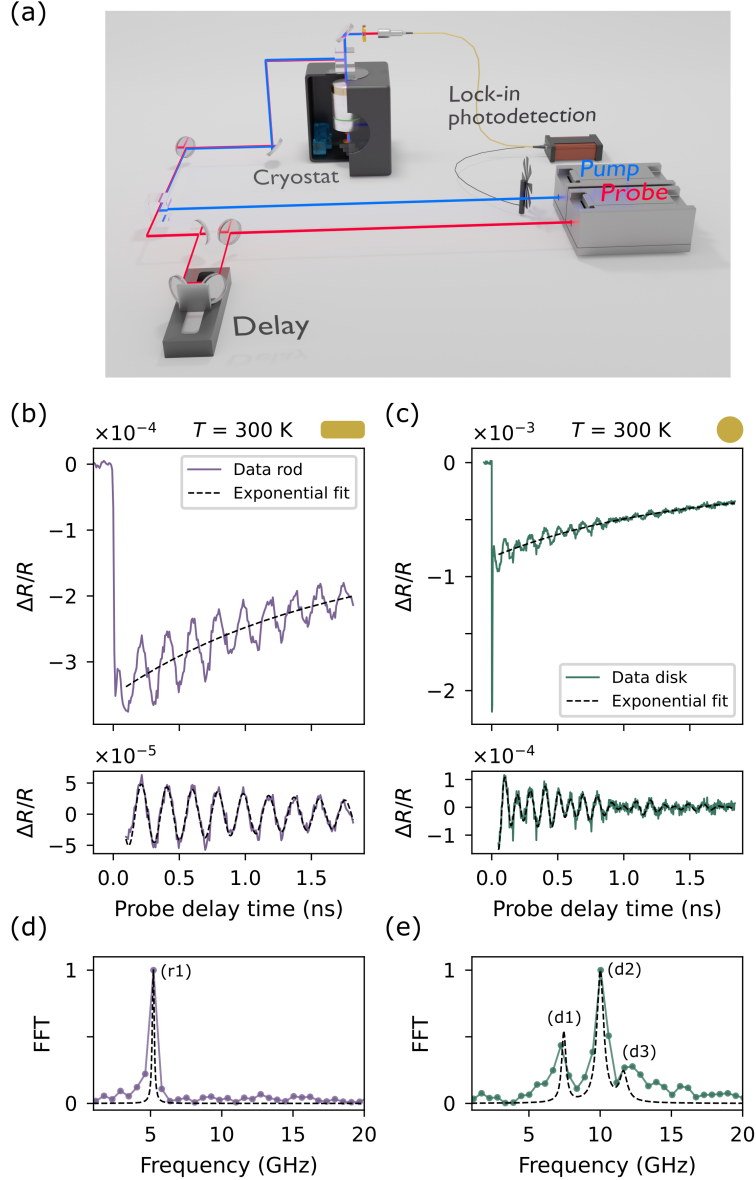

**Figure 2**. Experimental setup and transient reflection signals. (a) Schematic illustration of the nondegenerate pump-probe setup for coherent acoustic phonon detection within a closed-cycle cryostat. (b,c) Top: Experimental differential reflection signals ($\Delta R/R$) of a single gold nanorod (b) and nanodisk (c) with $L$ = 250 nm and $D$ = 140 nm, respectively, at room temperature ($T$ = 300 K). Bottom: same signal with the exponential cooling background subtracted. One or more damped sinusoid functions were fitted to the data (eq. 1). (d,e) Fast Fourier transform (FFT) of the phononic signals in panels (b,c), respectively, along with a Lorentzian multi-peak curve (dashed black line) displaying the frequency spectrum of the fitted curves in panels (b,c). A single oscillatory component (r1) was observed for the nanorod, while three modes (d1, d2, and d3) were distinguished for the disk.

The acoustic vibrations were modeled using continuum mechanics (valid down to 5 nm characteristic size [14]), which consisted of solving Navier's equation along with boundary conditions (see numerical simulation details in the Methods section). The average magnitude of the displacement field was computed to estimate the relative contribution of the different mechanical modes to the optically detectable change in reflection (Figure 3a-c). Figure 3a shows the average displacement of a gold nanorod along its long axis (*x*-axis direction) for different lengths, *L*. A strong branch corresponding to the extensional-like mode, labeled as (r1), is observed. The mechanical displacement of the nanodisks in the radial and *z*-direction are shown in Figure 3b and Figure 3c, respectively. Two main branches can be seen in Figure 3b, labeled as (d2) and (d3), of which the lower frequency corresponds to the main radial breathing-like mode. A third mode, with a smaller amplitude and displacement, mainly in the vertical direction, is observed in Figure 3c and labeled (d1). The deformation profiles corresponding to each mode for two representative nanorod and nanodisk sizes are displayed in Figure 3d,e. The measured frequencies at room temperature are plotted in Figure 3a-c as black circles and triangles. The experimental frequencies closely match the simulated ones, except for the vibrational mode (d1), which is only seen for the 140 nm diameter disks. This mode has a frequency of 7.4 GHz, higher than the expected 5 GHz. This discrepancy is likely due to differences in the mechanical coupling between the nanoparticles and the substrate in the simulated model compared to the actual systems.[26]

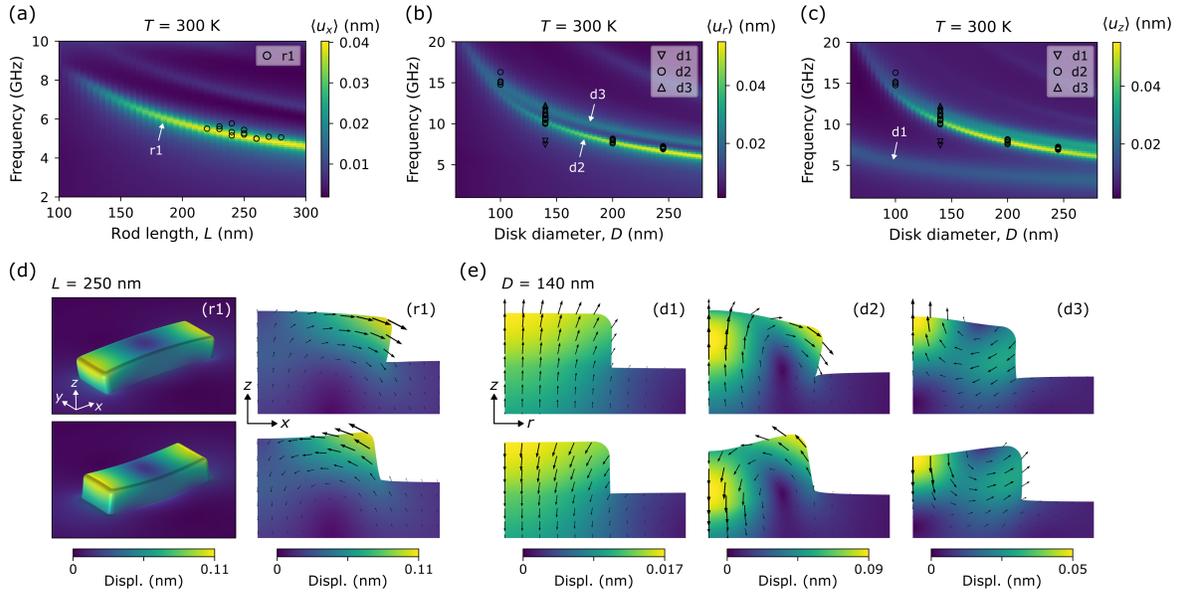

**Figure 3**. Mechanical vibrational modes at room temperature (*T* = 300 K). (a) Frequency-domain simulation of the average mechanical displacement of a gold nanorod along the *x*-direction, $\langle u_x \rangle$. The nanorod length varies between 100 nm and 300 nm; the width and height are fixed at *W* = 100 nm and $H_r$ = 45 nm, respectively. The most intense branch, indicated with an arrow, corresponds to the extensional-like mode, labeled as (r1). The measured frequencies are shown as black empty circles. (b,c) Simulation of the average displacement of a gold nanodisk in the radial direction, $\langle u_r \rangle$, and the *z*-direction (c), $\langle u_z \rangle$. The two main radial modes, labeled as (d2) and (d3), and a third mode (d1) with a displacement mostly in the *z*-direction, are marked with arrows. The experimental data for the different measured sizes are shown with black empty circles and triangles. (d,e) Calculated deformation geometries at the main mechanical resonances of a nanorod (d) with length *L* = 250 nm and a nanodisk (e) with diameter *D* = 140 nm, indicated in panels (a-c). A scale factor of 100× was applied to highlight the deformation. Arrows indicate the direction of the displacement field. An increase in the lattice temperature $\Delta T_L$ = 10 K and an isotropic loss factor $\eta$ = 0.1 were used in all simulations.

Figure 4a,b shows differential reflection traces (exponential background subtracted) of a single nanorod of length *L* = 250 nm and a single nanodisk of diameter *D* = 140 nm, at different temperatures. Starting at *T* = 300 K (or *T* = 350 K), the temperature was gradually lowered in 50 K steps until the lowest possible temperature of 4 K was reached. To account for the thermal contraction of the sample, the cryostat stage was allowed to settle for about 15-20 min between each cooling step. A clear increase in the resonant frequency is observed when the temperature decreases. In Figure 4c,d, the frequency shift with respect to its maximum value (occurring at low temperatures) is shown for each nanorod and nanodisk size. All the frequency-shift curves exhibit a linear temperature dependence at relatively high temperatures, converging to a constant value when approaching 4 K.

A semi-empirical formula for the temperature dependence of the elastic modulus, $E$, due to anharmonic effects of the lattice vibrations was proposed by Wachtman et al.[30]: $E(T) = E_0 - BTe^{-T_D/2T}$,[31,32] where $E_0$ is the zero-temperature Young's modulus, $B > 0$ is the slope of the Young's modulus-temperature curve at relatively high temperatures, and $T_D$ corresponds to the Debye temperature in the limit of high temperatures.[31] Now, assuming that the nanostructure-substrate mechanical coupling remains unaltered with changes in temperature, we can consider an Euler-Bernoulli model for the resonant frequency dependence with the Young's modulus: $f(E) \propto L^{-1}\sqrt{E/\rho}$,[33] where $L$ is the nanorod length, and $\rho$ the material mass density. Putting everything together, we obtain:

$$f^{(e)}(T) = f_0^{(e)}\left[1 - (B/E_0)Te^{-T_D/2T}\right]^{1/2} \qquad (2)$$

where $f_0^{(e)} = f^{(e)}(T = 0)$ is the frequency at the zero-temperature limit, and the superindex *(e)* indicates that this is a purely elastic material. In Figure 4c, we fit the average nanorod frequency data using this model, obtaining $T_D \simeq 168$ K, in excellent agreement with values reported in the literature for the Debye temperature of pure gold.[34–37] By using the former formula, we neglect the effects of thermal expansion on the size of the nanostructures. This assumption is reasonable because the observed increase in resonant frequencies with decreasing temperature is too large to be explained by thermal expansion. Indeed, the maximum variation in the nanorod length can be estimated as $\Delta L \sim \alpha_{Au}\bar{L}\Delta T_{max} \sim 1$ nm, where $\alpha_{Au} = 14.2 \times 10^{-6}$ K$^{-1}$ is the gold coefficient of linear thermal expansion,[38] $\bar{L} = 250$ nm the mean length of the measured structures, and $\Delta T_{max} = 296$ K, the largest temperature change. Such change in the nanoparticle length leads to a change in the resonant frequency $\Delta f \sim \Delta L/\bar{L} \sim 0.4\%$, which is one order of magnitude lower than the measured frequency shift of about 5% (Figure 4c,d).

It should be noted that the temperature on the horizontal axes of Figure 4c,d refers to the ambient temperature set by the cryostat. The transient nanoparticle temperature is expected to be slightly higher after absorbing an incident laser pulse. The lattice temperature increase, $\Delta T_L$, can be estimated as $\Delta T_L \simeq E_{abs}/mc_p$,[16] where $m$ is the nanoparticle mass, $c_p = 129$ J kg$^{-1}$K$^{-1}$ the specific heat capacity of gold,[38] and $E_{abs} = \sigma_{abs}(\lambda)E_{pump}/A_{pump}$ is the pump pulse energy absorbed by the nanoantenna, being $\sigma_{abs}$ its optical absorption cross-section, $E_{pump}$ the pump pulse energy and $A_{pump}$ the area of the laser spot. The calculated temperature increases, considering the average pump power $E_{pump} = 30$ µW and measured spot sizes (0.9 µm FWHM), fall in the range of 4 K to 7 K for all the measured nanostructures.

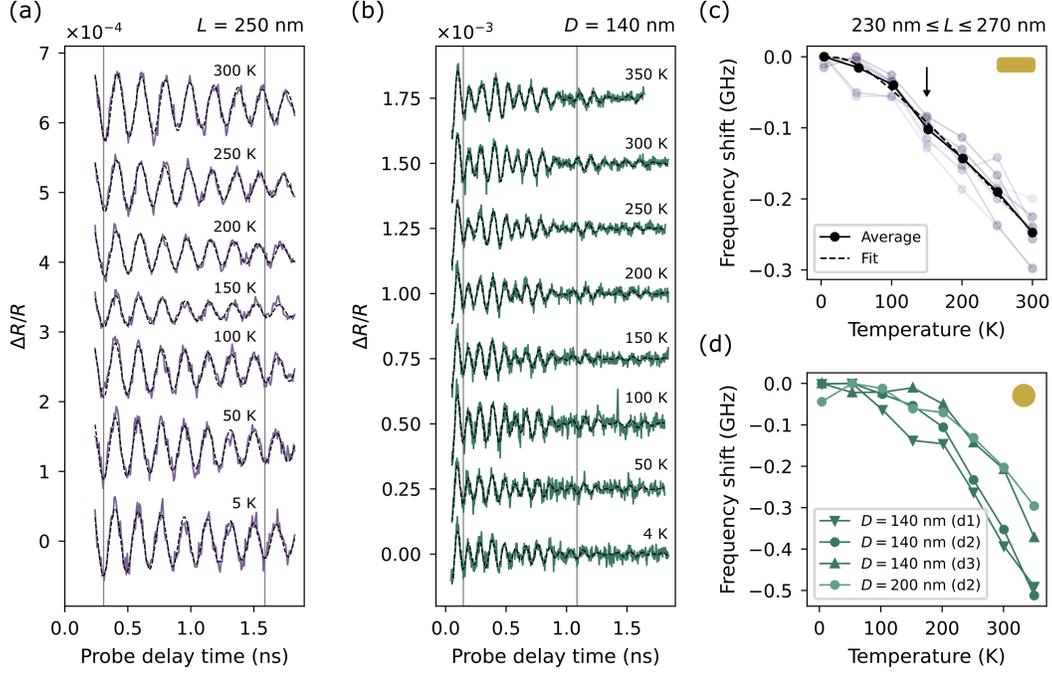

**Figure 4**. Temperature-dependent nanoparticle vibrational frequencies. (a,b) Experimental differential reflection traces ($\Delta R/R$) for a single gold nanorod of length $L$ = 250 nm (a) and a single gold nanodisk of diameter $D$ = 140 nm (b), at different temperatures indicated in the graphs. The vertical gray lines serve as guides for the eyes to observe the change in the vibration frequency. (c,d) Temperature dependent frequency shift measured on nanorods (c) and nanodisks (d) of different sizes. The average frequency-temperature curve of the nanorods (black circles) was fitted using eq. 2, obtaining $B/E_0 = 3.8 \times 10^{-4}\,\mathrm{K^{-1}}$, and $T_D$ = 168 K. An inflection point, indicated with an arrow in panel (c), is observed at ∼ 150 K.

The total $Q$-factor of the nanoresonator can be expressed in terms of the intrinsic dissipation mechanisms, also called internal friction, $Q_{\mathrm{int}}$, and the extrinsic contribution, $Q_{\mathrm{ext}}$, which is mainly related in this case to the radiation of acoustic energy to the substrate, as $Q^{-1} = Q_{\mathrm{int}}^{-1} + Q_{\mathrm{ext}}^{-1}$. Let us first consider an intrinsic loss mechanism according to the standard and very general Zener's model of anelasticity:[32,39,40]

$$Q_{\mathrm{z}}^{-1} = \Delta \frac{\omega\tau}{1+(\omega\tau)^2} \qquad (3)$$

where $\Delta$ is the relaxation strength (normalized change in a relevant mechanical modulus between relaxed and unrelaxed states), $\omega = 2\pi f$ is the angular frequency, and $\tau$ denotes the relaxation time. Eq. 3 exhibits a Lorentzian behavior with a peak dissipation at $\omega\tau = 1$, having a minimum quality factor value at this condition. This behavior can be qualitatively understood in the following way: if the vibration frequency $\omega$ is much smaller than the effective relaxation rate $\tau^{-1}$, the system remains in equilibrium, resulting in minimal energy dissipation. Conversely, if $\omega \gg \tau^{-1}$, the system does not have sufficient time to relax, leading to little energy dissipation as well. Significant dissipation occurs only when the vibration frequency is on the order of the effective relaxation rate. To take into account several possible relaxation mechanisms that might give rise to these dissipation peaks (known as *Debye peaks*), like point-defect relaxation, defect pair reorientation, dislocation relaxation, and boundary relaxation,[40] we need to make some assumptions. Indeed, for a very general thermally activated defect motion mechanism, we can say that $\tau$ obeys an Arrhenius equation of the form $\tau(T) = \tau_0 e^{E_b/k_b T}$. Then, assuming that $\Delta$ is independent of the temperature, which is equivalent to considering a negligible variation in the elastic modulus due to the action of a given dissipation mechanism, we obtain:

$$Q_z^{-1} = \Delta \frac{2\pi f \tau_0 e^{E_b/k_b T}}{1 + (2\pi f \tau_0)^2 e^{2E_b/k_b T}} \qquad (4)$$

where $f = f(T)$ is the temperature-dependent frequency, $\tau_0^{-1}$ is an attempt frequency and $E_b$ is the activation energy (or barrier height). Then, we fit our experimental data to the following equation:

$$Q^{-1}(T) = Q_z^{-1}(T) + Q_0^{-1} \qquad (5)$$

where $Q_0$ accounts for the remaining extrinsic and intrinsic dissipation mechanisms, which are assumed to be independent of temperature. The model described by eqs. 4 and 5 for the dependence of the $Q$-factor on temperature is illustrated in Figure 5a. As previously discussed, according to eq. 3, a dissipation peak should appear when the resonant condition $\omega\tau = 1$ is satisfied at a given temperature $T_m$:

$$\omega\tau = 1 = 2\pi f(T_m)\tau_0 e^{E_b/k_b T_m} \qquad (6)$$

Figure 5b shows the experimental average $Q$-factors obtained for the nanorods and nanodisks of different sizes, along with the fitting curves by eq. 5. A minimum of the $Q$-factor (*i.e.,* a dissipation peak) is clearly observed at $T_{m,r}$ = 149 K for the nanorods and $T_{m,d}$ = 168 K for the mode (d2) of the 140 nm diameter nanodisks. In addition, the relaxation theory predicts an inflection point in the frequency-temperature curve close to $T_m$ (see Supplementary Material, S1 for details), which has been observed in Figure 4c and marked with an arrow. Based on the fitting parameters (detailed in the caption of Figure 5), we calculate a $Q_z$ of 250 for the nanorods and 530 for the nanodisks at room temperature. Additionally, we estimate a $Q_z^{min}$ of 58 for the nanorods and 125 for the nanodisks at $T_m$. These numbers indicate that internal friction mechanisms are not the primary cause of the low measured values of the total $Q$-factor. In the work by Major et al.,[41] slightly lower intrinsic values (90 on average and 129 at the highest extreme) were estimated at room temperature for gold nanowires in a geometry minimizing substrate anchoring, whereas Wang et al. reported a value of about 200 for chemically synthesized gold nanoplates supported by a substrate designed to reduce acoustic radiation into it.[1] Accordingly, our data shows the role played by extrinsic mechanisms, such as anchoring to the substrate or mechanical energy dissipation through surface contact, included in $Q_0$. Indeed, as $Q_0 << Q_z$ ($Q_0$ = 23.9 for rods, 20.3 for 140 nm disks, and 14.8 for 200 nm disks), the result is a poorer performance of these plasmonic nanoresonators regarding the total $Q$-factor. Certainly, a dependence on the nanoresonator size and geometry is observed, obtaining higher values of the total $Q$-factor for the nanorods than for the nanodisks. Furthermore, there is a noticeable decrease in the average values with increasing disk diameter, following an enhanced transfer of acoustic energy into the substrate.

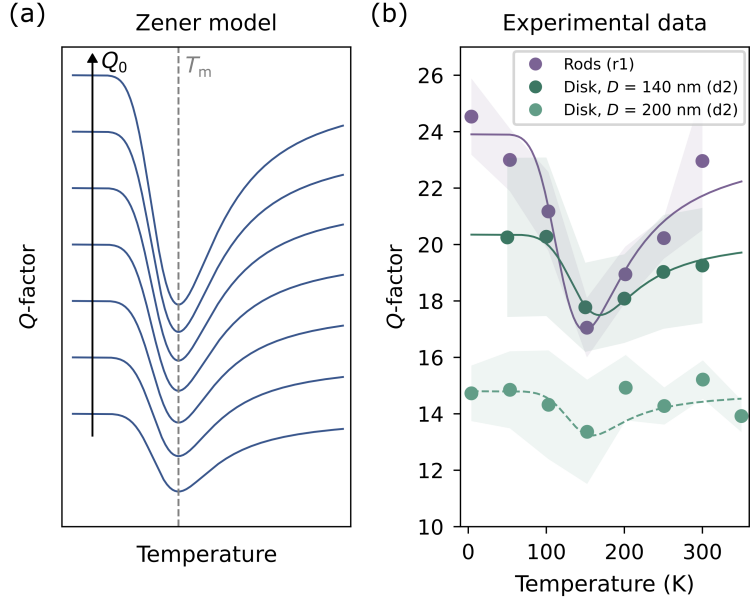

**Figure 5**. Temperature-dependent acoustic dissipation. (a) Illustration of the temperature-dependent $Q$-factor behavior according to Zener's model (eq. 5), where $Q_0$ represents a temperature-independent contribution to the total $Q$, and $T_m$ is the temperature of the dissipation peak (eq. 6). (b) Average measured $Q$-factors (circles) for the nanorods (mode r1) and nanodisks (mode d2) of different diameters, along with the fit curves by eq. 5 (solid lines). The shaded area around the experimental points represents the standard error of the mean. For the nanorods data, we obtained: $\Delta_r = 0.034$, $\tau_{0,r}^{-1} = 2.2$ THz, $E_{b,r} = 0.053$ eV, and $T_{m,r} = 149$ K, while for the 140 nm diameter disks, we obtained: $\Delta_d = 0.016$, $\tau_{0,d}^{-1} = 5.5$ THz, $E_{b,d} = 0.064$ eV, and $T_{m,d} = 168$ K. The dashed curve for the 200 nm diameter disks is the result of evaluating eq. 5 with the fitting parameters obtained for $D = 140$ nm, taking $Q_0 = 14.8$.

The activation energy values, $E_b$, obtained from the curve fits in Figure 5b (see caption), are close to the known activation energy of the Niblett-Wilks dissipation peak of about 0.05 eV.[42–44] Additionally, the computed attempt frequencies are in the order of 1 THz, as expected for a thermally activated motion of point defects in solids.[45] This subsidiary dislocation relaxation peak (apart from the main dissipation peak, known as the Bordoni peak, expected at much higher temperatures, not reached in this study) is associated with a geometrical kink-antikink nucleation process[46] and was first observed by Niblett and Wilks[42] in high-purity polycrystalline copper at vibration frequencies between 0.38 and 1.1 kHz, with an activation energy of 0.07 eV. Later on, an activation energy of 0.1 eV and an attempt frequency of 2 THz were deduced by Okuda for the same dissipation peak in pure gold from experiments with vibration frequencies between 1 Hz and 1 kHz.[47] For pure copper, an activation energy of 0.045 eV and an attempt frequency of 0.08 THz were obtained from measurements with vibration frequencies ranging from 1 Hz to 10 MHz.[47] Alnaser and Zein obtained an activation energy of 0.065 eV and an attempt frequency of about 2 THz for a low-purity polycrystalline aluminum sample at kHz and MHz frequencies.[48] Our results indicate that the observed dissipation peak might be explained by this mechanism, given the similarity in the values for activation energy and attempt frequency. However, this conclusion should be treated with caution, as these values could be influenced by the sample's manufacturing process and its polycrystalline or amorphous structure, such as grain size, density, and grain boundaries.

**Conclusions**

In conclusion, we studied the two main parameters of a mechanical plasmonic nanoresonator — its frequency and quality factor — as a function of temperature, for temperatures ranging from ambient to 4 K. Our findings highlight that external dissipation factor play a more crucial role in enhancing quality factors compared to internal friction. By modifying the nanoresonator's environment to manage these external losses, we can design structures with significantly higher vibrational frequencies and quality factors. In particular, we have investigated gold nanorods and nanodisks of different sizes fabricated by e-beam lithography, supported by solid substrates with a contact surface mediated by a thin adhesion layer. In all cases, an increase of the oscillation frequency with decreasing temperature was observed, showing a linear relationship and saturation when approaching liquid helium temperatures. This trend can be explained by considering the dependence of the elastic modulus on temperature through a semi-empirical formula involving the Debye temperature, which was estimated to be 168 K in this work. To describe the behavior of the quality factor, however, it was necessary to consider the nanostructure as an anelastic solid that presents a dissipation peak at around 150 K, associated with a thermally activated process. With this model, values of activation energies and attempt frequencies responsible for the thermal process were obtained. By comparing these results with the literature, it was identified that the Niblett-Wilks mechanism could explain our observations. Moreover, by fitting the experimental curve of $Q$ as a function of temperature with the proposed model we estimated values of $Q_z$ due to thermally dependent internal friction that are much higher than the total $Q$ usually measured in these systems. Therefore, to maximize $Q$, the dissipation due to external factors should be lowered by modifying the interaction of the nanoresonator with its environment, which is feasible since the limit imposed by internal factors is not so restrictive. However, we also found that the precise value of $Q_z$ is also determined by the geometry of the nanoresonator as well as the particular mode in which it vibrates. Likewise, we have observed that the vibrations in rod-type structures (mode (r1)) and disk-type structures (mode (d2)) exhibit different deformation fields and, consequently, different acoustic energy radiation into the substrate. As a final remark, our results allow for the comprehensive design of structures without losing the high vibration frequency while simultaneously controlling external losses with the aim of increasing the $Q$-factor.

**Methods**

Sample Fabrication
The gold nanostructures were fabricated by electron-beam lithography. First, the substrates were coated with a PMMA (Poly(methyl methacrylate)) resist on which the shape of the nanostructures was defined. For the nanorod sample, the electron beam-patterned resist was first covered with a 2 nm thick Cr adhesion layer and then with a 35 nm thick Au film by electron-beam evaporation. For the nanodisks sample, organic silane MPTMS ((3-mercaptopropyl)trimethoxysilane) was used as a molecular adhesion layer between the 45 nm thick gold and the borosilicate glass substrate. Both adhesion layers are expected to improve the adhesion of the metallic nanostructures to the supporting substrate. Lastly, acetone was used to remove the excess of resist and metal.

Numerical Simulations
Numerical calculations were performed using the Wave Optics and Structural Mechanics modules of the FEM solver software COMSOL Multiphysics. For the optical absorption simulations of the gold nanodisks, the effect of the thin ($< 1$ nm) molecular adhesion layer is expected to be negligible;

hence it was not considered. The acoustic response of the nanostructures was modeled using a continuum mechanics approach by solving Navier's equation along with boundary conditions in the frequency domain. An impulsive thermal strain, $\varepsilon_{\text{th}}$, proportional to the increase in the lattice temperature, $\Delta T_{\text{L}} \sim 10$ K, was considered in the metal domain as the displacive excitation mechanism that sets the particle in motion: $\varepsilon_{\text{th}} = \alpha \Delta T_{\text{L}}$, where $\alpha$ is the coefficient of linear thermal expansion. An isotropic loss factor $\eta = Q^{-1} = 0.1$ was implemented in the gold domain to account for the intrinsic dissipation mechanisms. Perfectly matched layers were used to truncate the substrate domain by absorbing the propagating acoustic waves traveling out of the computational domain. For further information on the mechanical simulations, see ref. [2].

**Supporting Information**
Modulus defect in the frequency-temperature curve. Reference values for properties of materials used in numerical simulations.

**Acknowledgments**
This work was partially funded by PICT 2021 IA 363 and PICT 2019 01886 (ANPCYT), PIP 112 202001 01465 (CONICET), UBACyT Proyecto 20020220200078BA. We acknowledge also funding and support from the Deutsche Forschungsgemeinschaft (DFG, German Research Foundation) under Germany's Excellence Strategy–EXC 2089/1–390776260 - e-conversion research cluster, the Bavarian program Solar Energies Go Hybrid (SolTech) and the Center for NanoScience (CeNS) at LMU Munich. A.V.B acknowledges funding support from the Alexander von Humboldt Foundation through the Georg Forster Award. T.P. thanks the Bavarian State Ministry of Science and Arts for support through the program EQAP. L.S. acknowledges funding support through a Humboldt Research Fellowship from the Alexander von Humboldt Foundation.

**Conflict of Interest**
The authors declare no competing financial interest.

**Supporting Information for:**

**Coherent Acoustic Phonons in Plasmonic Nanoparticles: Elastic Properties and Dissipation at Low Temperatures**

Hilario D. Boggiano,[1] Thomas Possmayer,[2] Luis Morguet,[2] Lin Nan,[2] Luca Sortino,[2] Stefan A. Maier,[3,4] Emiliano Cortés,[2] Gustavo Grinblat,[1,5] Andrea V. Bragas,*,[1,5] Leonardo de S. Menezes*,[2,6]

[1] Universidad de Buenos Aires, Facultad de Ciencias Exactas y Naturales, Departamento de Física. 1428 Buenos Aires, Argentina.
[2] Chair in Hybrid Nanosystems, Nanoinstitute Munich, Faculty of Physics, Ludwig-Maximilians-Universität München, 80539 München, Germany.
[3] School of Physics and Astronomy, Monash University, Clayton, Victoria 3800, Australia.
[4] Department of Physics, Imperial College London, London SW7 2AZ, U.K.
[5] CONICET - Universidad de Buenos Aires, Instituto de Física de Buenos Aires (IFIBA). 1428 Buenos Aires, Argentina.
[6] Departamento de Física, Universidade Federal de Pernambuco, 50670-901 Recife-PE, Brazil.

* Email for A.V.B.: bragas@df.uba.ar. Email for L. de S.M.: L.Menezes@physik.uni-muenchen.de

Content:

S1. Modulus defect in the frequency-temperature curve.
S2. Reference values for properties of materials used in numerical simulations.

**S1. Modulus defect in the frequency-temperature curve.**

The relaxation theory employed also predicts a change in the resonant frequency of a given vibrational mode with the temperature according to the relation:[1]

$$\frac{f^{(r)}-f_0^{(r)}}{f_0^{(r)}} = \frac{\Delta f^{(r)}}{f_0^{(r)}} = \frac{\Delta}{4}\left\{1 + \tanh\left[\ln\left(2\pi f^{(r)}\tau_0 e^{E_b/k_b T}\right)\right]\right\} \qquad (S1)$$

where $f_0^{(r)}$ is the relaxed value of the resonant frequency. When the latter variation is present, an inflection point (modulus defect) must be observed in the frequency-temperature curve at a temperature close to $T_m$. In Figure 4c, the expected inflection point is indicated with an arrow. In Figure S1, we fit the frequency-temperature average data for the nanorods to the full model consisting of a temperature-dependent elastic modulus (eq. 2) and a modulus defect due to a thermally activated relaxation effect (eq. S1).

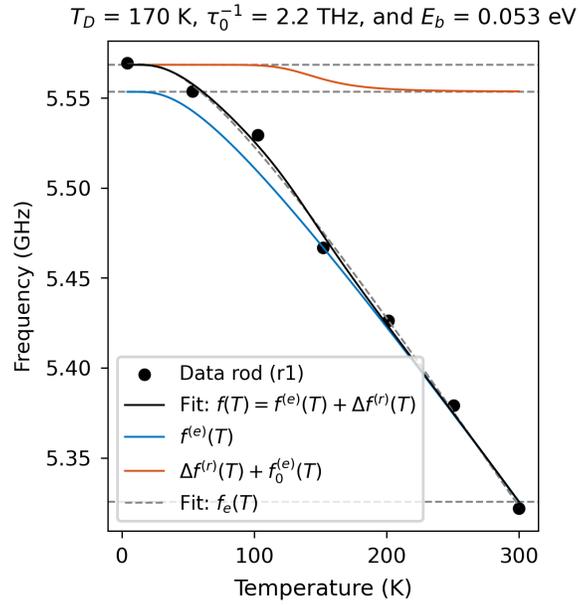

$T_D = 170$ K, $\tau_0^{-1} = 2.2$ THz, and $E_b = 0.053$ eV

**Figure S1**. Contributions to the temperature dependence of frequency due to temperature-dependent elastic modulus, $f^{(e)}(T)$, and a thermally activated relaxation process, $f^{(r)}(T)$.

## S2. Reference values for properties of materials used in numerical simulations.

| Material | $n$ | $k$ | $E$ (GPa) | $v$ | $\rho$ (kg m⁻³) | $\alpha$ (K⁻¹) |
|---|---|---|---|---|---|---|
| Au | Ref. [2] | Ref. [2] | 78 | 0.44 | 19300 | 14.2×10⁻⁶ |
| Cr | Ref. [2] | Ref. [2] | 279 | 0.21 | 7150 | |
| Borosilicate | 1.5 | 0 | 64 | 0.2 | 2400 | |

**Table S2**. Reference values for properties of materials.[3] $n$: refractive index, $k$: extinction coefficient, $E$: Young's modulus, $v$: Poisson's ratio, $\rho$: mass density, $\alpha$: coefficient of linear thermal expansion.